\documentclass[useAMS,usenatbib]{mn2e}

\usepackage{epsfig}

\def\lesssim{\mathrel{\hbox{\rlap{\hbox{\lower4pt\hbox{$\sim$}}}\hbox{$<$}}}}
\def\gtrsim{\mathrel{\hbox{\rlap{\hbox{\lower4pt\hbox{$\sim$}}}\hbox{$>$}}}}

\title[Predictions for Fomalhaut]
{Predictions for a planet just inside Fomalhaut's eccentric ring}

\author[Quillen]{Alice C. Quillen 
\\ 
Department of Physics and Astronomy, University of Rochester, Rochester, NY 14627; aquillen@pas.rochester.edu}

\begin{document}
\label{firstpage}

\maketitle

\begin{abstract}
We propose that the eccentricity and sharpness of the edge of Fomalhaut's 
disk are due to a planet just interior to the ring edge.
The collision timescale consistent with the disk opacity is long
enough that spiral density waves cannot be driven near the planet.
The ring edge is likely to be located at the boundary of a chaotic zone
in the corotation region of the planet.  We find that this zone can
open  a gap in a particle disk as long as the collision
timescale exceeds the removal or ejection timescale in the zone.
We use the slope measured from
the ring edge surface brightness profile to
place an upper limit on the planet mass. 
The removal
timescale in the chaotic zone is used to estimate a lower limit.
The ring edge has eccentricity caused  
by secular perturbations from the planet.  
These arguments imply
that the planet has a mass between that of Neptune and that of Saturn, 
a semi-major axis of approximately 119 AU and 
longitude of periastron 
and eccentricity, 0.1, the same as that of the ring edge.

\end{abstract}


\section{Introduction}

The nearby star Fomalhaut 
hosts a ring of circumstellar material \citep{aumann85,gillett85}
residing between 120 and 160 AU from the star 
\citep{holland98,dent00,holland03}.  
The ring is not axisymmetric.
{\it Spitzer Space Telescope} infrared 
observations of Fomalhaut reveal a strong brightness asymmetry
in the ring \citep{stapelfeldt04,marsh05}.  Submillimeter observations 
are less asymmetric in brightness but also imply that the ring is
offset, with the southern side nearer the star than the opposite side 
\citep{holland03,marsh05}.  
Recent {\it Hubble Space Telescope} ({\it HST}) observations show
that this ring has both a steep 
and eccentric inner edge \citep{kalas05}. 
In this paper we explore
dynamical scenarios involving a planet that can 
account for both the eccentricity
of the ring edge and its sharp or steep surface brightness edge profile.

Two classes of theoretical models exist for non-transient eccentric rings that
do not rely on dynamics induced by radiation pressure.
These are the pericenter glow model \citep{wyatt99} 
and the self-gravitating eccentric ring models (e.g., 
\citealt{goldreich79,tremaine01,papaloizou05}).
The self-gravitating ring models have primarily
been used to explain eccentric planetary rings.
%The apsidal alignment in the ring is maintained
%by self-gravity, overcoming strong differential precession
%due to the quadrupolar gravitational field of the planet.  
Though the structure of the ring edge can impact the models 
\citep{chiang00}, the ring edges
are not integral to the model, instead the rings are truncated by
torques driven by neighboring satellites.

%Neverthless the structure of the ring edge can impact the models.
%At the disk edge, ring self-gravity, interparticle collisions, 
%and differential precession balance each other to achieve equilibrium
%\citep{chiang00}.
%
%Eccentric ring models have primarily been applied to planetary systems with
%strong differential precession. On the other hand, circumstellar disks  such
%as that hosted by Fomalhaut,
%are likely to be nearly Keplerian.   \citet{tremaine01} has shown
%that Keplerian disks can support lopsided ($m=1$) slow modes. 
%This study predicts the spectrum of modes given specific density profiles,
%including Gaussian rings.
%As is true for the eccentric planetary ring models, 
%the disk edge must be caused or maintained by a different dynamical process. 

The pericenter glow model 
can account for eccentricity in the disk surface brightness distribution
of Fomalhaut's disk \citep{stapelfeldt04,marsh05}.
Secular perturbations from a planet interior to the ring cause
particle eccentricities to be coupled  with their longitudes of periastron.
The forced particle eccentricities cause an asymmetry in the dust distribution
such that the ring periastron is aligned with the planet's periastron.
Previous studies have not placed constraints on the location of
the planet causing the forced eccentricity in Fomalhaut's disk, 
consequently constraints on the planet's mass and 
eccentricity are lacking \citep{wyatt99,marsh05}.
%By requiring the age of the system to exceed the precession timescale 
%in the ring \citet{marsh05} suggested that the planet responsible
%for the eccentricity of Fomalhaut's ring is  
%greater than an Earth mass. 
%\citet{wyatt05} has
%explored transient models where this condition is violated. 

%These models most often predict that the eccentricity gradient is positive
%in the ring.

%To place better constraints on a hypothetical planet residing within the disk edge 
We briefly review the observed properties of Fomalhaut's disk.
The recent {\it HST} observations have revealed that
the ring edge has eccentricity $e_{edge} = 0.11 \pm 0.01$,
periastron at $PA = 170^\circ$, inclination $65.6^\circ$,
and a semi-major axis $a_{edge} = 133$~AU  \citep{kalas05}.
The surface brightness at the edge drops by 
a factor of 2  within 10 AU (see Figure 3 by \citealt{kalas05}).
This can be compared to
the resolution of {\it HST}, $0.1''$, corresponding 
to only 0.75~AU at the distance of Fomalhaut. 
The slope of the disk edge was modeled with 
a disk scale height of 3.5AU corresponding
to an opening angle of $1.5^\circ$ \citep{kalas05}.
However, the observed disk edge slope could either be due to the thickness
of the disk or a drop in the planar surface density profile.
Assuming an exponential 
dust density distribution in the form 
$\exp{\left(-{\left(a_{edge}-a\right)\over h_r} -{|z|\over h_z}\right) }$,
the observed disk edge slope implies that either
$h_r/r \sim 0.026$ and $2 h_z < h_r$ or the disk
aspect ratio $h_z/r \sim 0.013$ and $2 h_z >  h_r$.
The equation of
hydrostatic equilibrium can be used to place a limit
on the velocity dispersion, $u$, of the dust particles,
${u\over  na} \lesssim 0.013$
where $n$ is the mean motion at a semi-major axis $a$.

The age of the star is $200 \pm 100$~Myr \citep{barrado}.  
The mass of the star is  $2 M_\odot$ \citep{song01}, so
the rotation period at 130~AU is 1000 years.
% see wyatt02 for luminosity and temp ref
This orbital rotation period divided by the star's age is $10^5$.  
The optical depth (normal to the disk plane) just interior to the ring edge
at $24\mu$m is $\tau \sim 1.6\times 10^{-3}$ \citep{marsh05}.
The collision time in the ring , $t_{col} \sim (\tau n)^{-1}$,
is a million  years or 1000 orbits. 
% note factor of 2 pi missed previously
% actually t_col = 2 pi Period /tau but many people add in another 1/3 here
Since this timescale is short, we can exclude Poynting-Robertson driven
resonance capture models for the dust, as argued
in detail by \citet{wyatt05}. 
%A model where an outward migrating planet
%captured planetesimals that then produce the dust through collisions
%\citep{wyatt06} could be explored were future observations
%of Fomalhaut
%to show a dust morphology more complex than a smooth
%eccentric ring.    

\section{The Pericenter Glow Model and an eccentric edge in Fomalhaut's disk}

We follow the theory for secular perturbations 
induced by a planet (e.g., \citealt{M+D,wyatt99}).
Secular perturbations in the plane can be described in terms of the
complex eccentricity variable,
%\begin{equation}
$z = e \exp{(i \varpi)}$,
%\end{equation}
where $e$ is the object's eccentricity and $\varpi$ is its longitude
of periastron (e.g., \citealt{M+D,wyatt99}).
The time variation of $z$ is
\begin{equation}
\dot z = z_{forced} + z_{proper}(t)
\end{equation}
where
\begin{equation}
z_{forced} = { b^2_{3/2}(\alpha) \over  b^1_{3/2}(\alpha) } 
e_p \exp{(i \varpi_p)}
\label{eqn:zforced}
\end{equation}
\citep{M+D,wyatt99}.
We denote the planet's semi-major axis, eccentricity 
and longitude of periastron as $a_p$, $e_p$, and $\varpi_p$,
respectively.
Here $\alpha = a_p/a$ if $a_p < a$ otherwise $\alpha = a/a_p$.  
The functions, $b^j_s(\alpha)$, are Laplace coefficients (see \citealt{M+D}
for definitions and numerical expressions).
If the planet's periastron does not vary (as for the two body
problem) then $z_{forced}$ is a constant of motion. 
In this case the forced complex eccentricity 
depends only the planet's semi-major axis and 
eccentricity, and not on its mass.
%Additional planets in the system (and the mass of the ring itself) would
%induce secular variations in $z_{forced}$.

The ratio of Laplace coefficients, 
$b^2_{3/2}(\alpha)/b^1_{3/2}(\alpha) < 1$,
%and is not a strong function of $\alpha$.
so the amplitude of the complex eccentricity variable,
$|z_{forced}|$, cannot exceed the planet's eccentricity.
If the planet is near the ring edge then $\alpha$ is near 1.
Near the planet 
$\lim_{\alpha \to 1} \left[{ b^2_{3/2}(\alpha)/  b^1_{3/2}(\alpha)}\right] = 1$
and $|z_{forced}|=e_p$.  If the planet is near the ring
edge then the forced eccentricity is equal to that of the planet.

We now consider the density distribution  from a distribution
of particles.
Particles with the same semi-major axis, different
mean anomalies
and zero free or proper eccentricities would be located along a single
ellipse.  If the free eccentricities are non zero then the density
distribution is smoother than the zero free eccentricity ellipse 
and has a width twice the free eccentricity multiplied by the
semi-major axis.
Consequently the observed steepness of the disk edge limits
the distribution of free eccentricities in the edge.
We denote the free eccentricity dispersion, 
$ u^2_e = {\left<e^2_{proper}\right>}$.
The slope of Fomalhaut's disk edge  $h_r/r \lesssim 0.026$ so
the free eccentricities in the disk edge are 
$u_e  \lesssim 0.026$.
If the planet is responsible for truncating the disk and 
limiting the distribution of free eccentricities then
we suspect that the planet is located near the disk edge 
and $\alpha$ is almost $1$.
%Equation~(\ref{eqn:zforced}) implies that the eccentricity
%at the ring edge is approximately equal to the planet's eccentricity.
If the ring eccentricity is due to secular perturbations
from a planet then the hypothetical 
planet's eccentricity $e_p$ is equal to
that of the edge or $e_p\sim 0.11$.

Since they are inelastic, collisions damp the eccentricities
and inclinations of an ensemble of particles.  This damping leads
to a distribution following nearly closed (non-self-intersecting) orbits.  
Near a planet the only non-intersecting
closed orbits consist of those with zero free eccentricity 
and with eccentricity equal to the forced eccentricity.
%The sharp edge is consistent with the short collision timescale
%and the forced eccentricity explanation for the ring's eccentricity.

\section{Spiral density waves and the collision timescale}

In a high opacity, $\tau \sim 1$, disk
spiral density waves are driven by a planet or satellite near
the planet.  A gap opens if the torque from the planet
exceeds that from accretion and the minimum gap
width is twice the size of the planet's Hill sphere 
(e.g., \citealt{borderies89}).     
%These limits can be used to place constraints on a hypothetical planet 
%maintaining a disk edge 
%(e.g., as done by \citealt{cokutau4} for CoKuTau4).

As pointed out two hundred years ago
by Poisson, some form of interaction between particles
is needed for secular transport to occur.  Satellites or planets do 
not exert a torque on a collisionless disk.
However, planetesimals in the corotation region 
are efficiently pumped to high eccentricity and ejected by the planet
or other interior planets (e.g., \citealt{david03,mudryk06}).  
In this case the width of
a gap opened near the planet would be given by Equation~(\ref{eqn:daz}),
as is discussed further in section 4.
For planet mass objects the width of this chaotic zone exceeds
that set by the Hill radius because 2/7 is smaller than 1/3.

The separation between collisional and collisionless disks is
important as the opacity of the disk (setting the collision
timescale) is an observable.
\citet{franklin80,goldreich80,lissauer98} showed 
that spiral density waves were efficiently
driven at a Lindblad resonance by a satellite 
when the collision timescale was above a critical one, $t_{crit}$, where
$t_{crit} \propto \mu^{-2/3}$, and 
$\mu \equiv {m_p \over M_*}$ is the ratio
of the planet mass divided by that of the star.
This has been confirmed
numerically with simulations of low opacity
collisional particle disks at individual Lindblad
resonances \citep{franklin80,hanninen92,espresate01}.
\citet{lissauer98} predicted this scaling
by comparing the period of excited
epicyclic oscillations at a Lindblad resonance with the collision timescale.
Near a planet a series of resonances is encountered.
The $j:j-1$ mean motion resonance (corresponding to the $m=j-1$ Lindblad
resonance)
has a period approximately equal to the renormalization factor in
Equation~(7) by \citet{quillen06} or 
\begin{equation}
p_e  \sim n^{-1} |{\delta_{1,0} a'} |^{-2/3} =
     n^{-1} \left|{3 j^2 \mu \sqrt{2} \over 2 \pi da}\right|^{-2/3}
\end{equation}
with coefficients described by this work, and evaluated above in 
the limit of large $j$.
In the limit of small $da$, and setting the critical timescale
to this period, $t_{crit}  = p_e$,
\begin{equation}
t_{crit} n \sim  \mu^{-2/3} j^{-2} \sim \mu^{-2/3} da^{2}.
\label{eqn:tcrit}
\end{equation}
%Note $1-\alpha = da \sim j^{-1}$.
We have recovered the scaling with planet
mass predicted by previous
works \citep{goldreich80,franklin80,lissauer98}
but have also included a dependence on distance from the planet.
%Our scaling for $da$ disagrees with Equation~(106)
%by \citet{goldreich80} that is consistent with
%$t_{crit} n  \sim \mu^{-2/3} j^{-4/3}$.
%However, this difference does not significant change our subsequent conclusions.

%When the oscillation period exceeds the collision time or
%\begin{equation}
%p_e > t_{col}
%\end{equation}
%then torques can be driven by planet or satellite induced spiral density
%waves. 

The above critical timescale, $t_{crit}$, (appropriate for small $da$)
increases with distance from the planet.
For a disk with a particular
collision timescale, spiral density
waves would be driven past a particular
distance from the planet.
Because the Hill sphere radius is proportional to the planet
mass to the 1/3 power,  Equation~(\ref{eqn:tcrit}) implies that
$t_{crit}$ is of order 1 at the Hill sphere radius. 
Only collisional disks, $\tau \sim 1$, could have a disk edge 
extending to the planet's Hill sphere.
%At the chaos zone boundary  
%$t_{crit} \propto \mu^{-2/21}$, only slightly larger
%than 1 for planet sized objects.
%We find that spiral density waves can't be driven within
%a chaos zone unless the disk opacity is of order 1.
Since the opacity of Fomalhaut's disk is sufficiently low
that spiral density waves cannot be driven into the 
disk by a nearby planet, the disk edge must be maintained
by a different dynamical process.

\section{Velocity dispersion at the edge of the chaos zone}

%Previous studies have identified 
%stability regions in collisionless disks. %(e.g., \citealt{barnes04,holman99}).
%Regions of instability in collisionless disks, even in multiple planet systems,
%are caused by the overlap of resonances \citep{lepage04}.
There is an abrupt change in dynamics as a function
of semi-major axis at the boundary of the chaos zone in the
corotation region near a planet.  The width of this
zone has been measured numerically and predicted theoretically
for a planet in a circular orbit  by predicting the semi-major
axis at which the first order mean motion resonances overlap
%using a mean motion resonance overlap criterion 
\citep{wisdom80,duncan89,murray97,mudryk06}.
The zone boundary is at
\begin{equation}
da_z \sim 1.3 \mu^{2/7}
\label{eqn:daz}
\end{equation}
%\citep{wisdom80,duncan89},
%where $\mu \equiv {m_p \over M_*}$ is the ratio
%of the planet mass divided by that of the star.
where $da_z$ is the difference between the zone edge semi-major axis
and that of the planet divided by the semi-major axis of the planet.

In section 2 we found that the free or proper eccentricities
are likely to be limited by the observed disk edge slope.  
A collision could convert a planar motion to a vertical 
motion similar in size, suggesting that the disk velocity
distribution is not highly anisotropic  
so we may assume $h_z/r \gtrsim  u_e $.
A collision could also increase or decrease
the particle semi-major axis and eccentricity. 
Particle lifetime is likely to be strongly dependent on semi-major axis,
so we expect a sharp boundary in the semi-major axis distribution.  
The slope in the disk edge is likely to be set by the vertical
scale height and velocity dispersion in the disk edge.
We estimate that $h_z/r \sim u_e \sim 0.013$.  Here the value of 0.013
is half the scale height measured by \citet{kalas05} (see discussion
at the end of section 2).

\begin{figure}
%\plotone{lifes.eps}
\includegraphics[angle=0,width=3.5in]{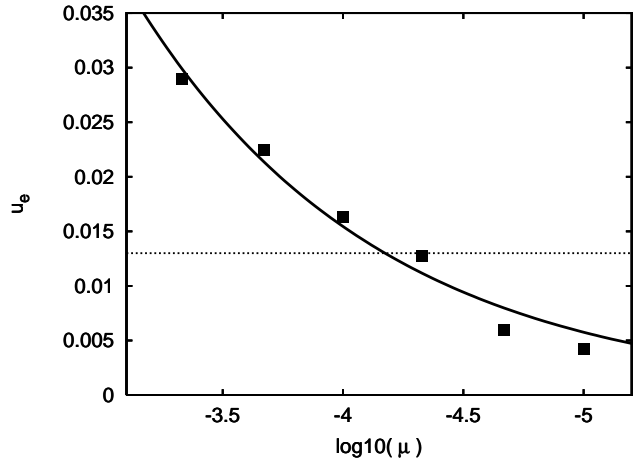}
\caption{
Eccentricity dispersion in the disk edge vs planet mass 
for a planet %in a circular orbit (circular points) and a planet
with eccentricity $e_p = 0.1$ (square points).
The solid line is $u_e = 0.8 \mu^{3/7}$.  The
scaling with $\mu^{3/7}$ is predicted for the libration in
the first order mean motion resonance just outside the corotation region.
The horizontal line shows the limit set from the observed
disk edge slope.
\label{fig:ee}
}
\end{figure}

Outside the chaos zone, planetesimals still experience perturbations from
the planet.  
These perturbations have a characteristic size
set by size of perturbations in the nearest mean-motion 
resonance that is not wide enough to overlap others and so is not part
of the chaotic zone.
Since particles in the edge reside outside
the chaotic zone, the velocity dispersion does not 
increase with time.  
Via numerical integration we find a relation, shown in 
Figure \ref{fig:ee}, between
the planet mass and the proper eccentricity dispersion 
just outside the chaos zone. 
The numerical integrations were carried out in the plane, 
using massless and collisionless particles under the gravitational
influence of only the star and a 
massive planet with eccentricity $e_p = 0.1$.
The initial particle eccentricities were set to
the forced eccentricity and the
and longitudes of periastron were chosen to be identical to that of 
the planet.  
Initial mean anomalies were randomly chosen.
The free eccentricity distribution was measured
after $10^5$ planetary orbits. However $u_e$ reached
the steady state value much earlier in the integrations, at a time
less a hundred planetary orbits and so shorter than the collision timescale. 

The libration width for a $j:j-1$ mean motion resonance has
$e^2 \sim \mu^{2/3} j^{2/3}$ (using the high $j$ limit
of equation 7 by \citealt{quillen06}).  
Setting $j \sim \mu^{-2/7}$ corresponding
to the chaos zone boundary we estimate an eccentricity dispersion
$u_e \sim \mu^{3/7}$ just outside the chaotic zone.  This dependence
on $\mu$
is shown as a solid line in Figure \ref{fig:ee} and is a good match
to the numerically measured dispersion values in the disk edge.

%Figure \ref{fig:ee} shows the eccentricity predicted for libration in
%a mean motion resonance just outside the chaos zone.

Fomalhaut's disk edge slope can be used to place a limit on the
planet mass if we assume that the disk edge is bounded by
the planet's chaotic zone. 
We use the limit, $u_e \sim 0.013$, based on the disk edge
slope, shown as
a horizontal line on Figure \ref{fig:ee}
to estimate the mass of the planet. This horizontal line is consistent
with the eccentricity dispersion at the edge of a chaos zone for a planet of 
mass $\mu \sim 7 \times 10^{-5}$.
As the mass of Fomalhaut is twice that of the Sun this corresponds
to a planet mass similar to that of Neptune.  
For this simulation the distance between
the planet semi-major axis and disk edge has
$da \sim 0.13$, approximately consistent with the $\mu^{2/7}$ law and 
corresponding to a planet's semi-major axis of 119 AU.

If the velocity dispersion in the disk edge is due to perturbations
from massive objects in the ring then it would exceed that
estimated from our integrations.  In this case the planet
maintaining the disk edge could be lower, but not
higher, than that estimated above.  
The mass ratio $\mu = 7 \times 10^{-5}$
can be regarded as an approximate upper limit for the planet mass.

%For $da = 0.13$ the planet's semi-major axis would be 119 AU.
%At this separation, the secular precession rate in the ring
%is only $\sim 2000$ planetary orbital
%periods. Secular perturbations from
%inner planets can cause the planet's longitude of periastron to vary, however
%the timescale for the planet to precess is likely to be longer than
%10^3 orbital periods (that appropriate for a modestly distant Jupiter
%mass planet).
%Because the collisional timescale is likely to exceed the planet's 
%precession rate collisions are likely to maintain
%the alignment between the ring and planet's periastron.

\section{Removal timescale from the corotation region}

%Previous studies have identified 
%stability regions in collisionless disks. %(e.g., \citealt{barnes04,holman99}).
%Regions of instability, even in multiple planet systems,
%are caused by resonances \citep{lepage04}.
%There is an abrupt change in dynamics as a function
%of semi-major axis at the boundary of the chaos zone in the
%corotation region, where mean motion resonances overlap.

Since they are inelastic, collisions damp the eccentricities
and inclinations of an ensemble, unless they are rapidly transported
elsewhere.     Since they
change orbital parameters, collisions cause diffusive spreading 
of the particle distribution in an initially sharp
ring edge.  A particle that is knocked
into an orbit with a semi-major axis within the chaotic zone can be 
scattered by the planet and ejected from the region.  
To maintain the low dust density within the ring edge, we infer
that the removal timescale within the ring must be shorter
than the rate that particles are placed interior to the ring.

We can approximate the dynamics with a diffusion equation where
diffusion due to collisions in the disk edge is balanced by the rapid removal
of particles on a timescale $t_{removal}$ inside the edge.
In steady state the diffusion equation 
\begin{equation}
{\partial  \over \partial a} \left(D {\partial N \over \partial a} \right)
\approx {N \over t_{removal}}
\label{eqn:diffusion}
\end{equation}
\citep{melrose80,varvoglis96},
where $N(a)$ is the number density of particles with semi-major axis $a$.
The diffusion coefficient, $D$,  depends on the collision time and
the velocity dispersion, $u$, in the disk, 
$D \sim { u^2 \over t_{col} n^2}$.  This diffusion
coefficient is similar to a viscosity and can be estimated
by considering the mean free path and particle
velocity differences set by the epicyclic amplitude.
The removal timescale $t_{removal}$ is set by the dynamics
within the chaos zone and depends on the planet mass and eccentricity.
The above equation is satisfied when $N(a)$ decays exponentially with a 
scale length $l$, and $l^2  = { D t_{removal}}$.

As the removal timescale depends on the planet mass and eccentricity
it is useful to write
\begin{equation}
t_{removal} = l^2/D
 =  \left({l \over h} \right)^2 t_{col} 
\label{eqn:remove}
\end{equation}
In section 4 we argued that the velocity distribution
is unlikely to be extremely anisotropic near the disk edge
and that this dispersion is set by the planet and the distance
to the disk edge.  Therefore we expect that  $l \sim h_z$.
Equation~(\ref{eqn:remove}) implies that 
in order for a planet to open a gap in a low opacity disk
it must be massive and eccentric enough that 
the removal timescale in the chaos zone exceeds the collision timescale.

\begin{figure}
%\plotone{lifes.eps}
\includegraphics[angle=0,width=3.5in]{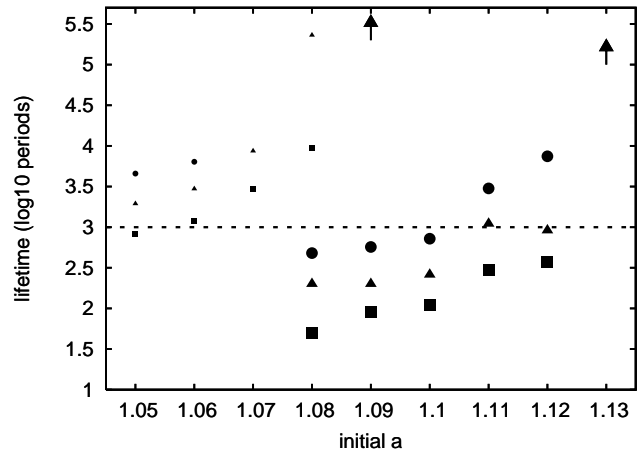}
\caption{
Removal timescale as a function of initial semi-major axis for
a planet mass $\mu = 10^{-4}$ (large points)
and $\mu = 2 \times 10^{-5}$ (small points).
%For both cases the planet eccentricity $e_p=0.1$.
%For each initial semi-major axis (shown on the $x$-axis),
%100 particles were integrated in the plane
%with initial eccentricities and longitudes
%of periastron identical to those of the planet, and randomly chosen  
%mean anomalies.
Particles were
removed from the integration when their eccentricity became larger than 0.5.
Squares, triangles and circles 
show the timescale when fewer than 
75\%, 50\% and 25\% of the particles remained in the integration,
respectively. 
%The $y$-axis shows log10 of the removal time in planetary orbital periods.
For an initial semi-major axis above or equal to 1.13, 
particles were not removed in less
than $10^5$ orbital periods (shown as the arrow on the upper right)
for $\mu=10^{-4}$ and for semi-major axis above or equal to 1.09
in less than $2 \times 10^5$ orbital periods for $\mu=2 \times 10^{-5}$ 
(shown as the arrow on the top middle).
The horizontal line shows the limit set from the 
collision timescale, $10^3$ orbits, estimated from Fomalhaut's disk opacity. 
To account for the absence of dust within the ring edge,
particle lifetimes within the chaotic zone
must be shorter than the collision timescale.
\label{fig:life}
}
\end{figure}

Previous works estimating ejection timescales in the
corotation region have primarily concentrated
on more massive mass ratios than $\mu = 10^{-4}$ \citep{david03,mudryk06}.
Consequently we have estimated this timescale from numerical integrations.
100 particles were integrated in the plane
with initial eccentricities and longitudes
of periastron identical to those of the planet, random  mean anomalies
and differing initial semi-major axes.
Particles were
removed from the integration when their eccentricity was larger than 0.5.
Figure \ref{fig:life} shows this removal timescale as a 
function of semi-major axis for planet mass
ratios $\mu=10^{-4}$, $ 2 \times 10^{-5}$ and eccentricity $e_p=0.1$.
Figure \ref{fig:life} shows that the removal timescale in
the chaotic zone for $\mu = 10^{-4}$ is similar or below to 
the estimated collision timescale for Fomalhaut, $10^{3}$ orbits, whereas
the removal timescale is longer than than this time
for $\mu \sim 2 \times 10^{-5}$.  In section 4
we estimated that the planet mass must be lower than $7\times 10^{-5}$.
Here we find that if the planet mass is below $\mu \sim 2 \times 10^{-5}$
then the chaotic zone would not be able to open a gap in Fomalhaut's 
particle disk.

The diffusion equation (Equation~\ref{eqn:diffusion}) neglects any
dependence of the diffusion coefficient or removal time on particle
radius  or eccentricity.    We also have not considered the role
of a particle size distribution and destructive collisions. 
The low scale height implied from the sharp edge implies
that fewer collisions are destructive than previously estimated (e.g, 
by \citealt{wyatt02}).  
A more sophisticated model is needed to more accurately predict the
edge profile as a function of planet mass, eccentricity,
collision timescale and particle size.

\section{Summary and Discussion}

We find that a planet accounting 
for both the eccentricity and edge of Fomalhaut's disk
is likely to have eccentricity similar to that of the disk edge 
or $e_p \sim 0.1$.  
Here we have assumed that the eccentricity of the ring is a forced
eccentricity due to the planet.  The sharp disk edge limits the free 
eccentricities in the ring edge, so the ring eccentricity equals the
forced eccentricity.  For a planet close enough to truncate
the ring, the forced eccentricity is approximately the same as
the planet eccentricity.

For high opacity or collisional disks ($\tau \sim 1$), 
a gap is only formed if the planet driven spiral
density waves can overcome the torque from
accretion.  A planet just large enough to open a gap will open one 
approximately twice 
the size of its Hill radius.   However, a collisionless disk can open
a larger gap, the size of the chaos  zone in the planet's corotation region.
%In general this zone is larger than the width of the planet's Hill sphere.
%
%
We find that spiral density waves can only be driven into
a disk within a chaotic zone if the disk opacity is of order 1.  
Fomalhaut's disk opacity, $\tau \sim 1.6 \times 10^{-3}$ \citep{marsh05}, 
is sufficiently low that spiral
density waves cannot be driven near the planet.
For low opacity disks, $\tau \lesssim  0.1$, 
a planet will
open a gap to the chaos zone boundary only if the collision
timescale exceeds the timescale for removal of particles
within the chaos zone.
We use this limit 
and numerical integrations to infer that
a mass of a planet sufficiently large to account for the sharp
edge in Fomalhaut's disk edge has  mass ratio
$\mu \gtrsim  2 \times 10^{-5}$.

The planet mass can be estimated from
the observed slope in the disk edge by assuming that the
ring edge is located at the edge of the planet's chaos zone
and the velocity dispersion at the ring edge is set by
resonant perturbations caused by the planet.
If the velocity dispersion estimated at the ring edge is
due to perturbations caused by bodies in the ring then the planet
mass must be lower than this estimate.  
This limits the planet mass ratio $\mu \lesssim 7 \times 10^{-5}$.

Our exploration suggests that there is a planet located
just interior  to Fomalhaut's ring with
semi-major axis $\sim 119 AU$, mass ratio 
$2 \times 10^{-5} \lesssim  \mu  \lesssim  7 \times 10^{-5}$ (corresponding
to between a Neptune and Saturn mass), and
and longitude of periastron and eccentricity,  $e_p\sim 0.1$,
the same as that of the ring edge.  
%The ring outer edge may be located
%at the  planet's 3:2 mean-motion resonance.
Arguments  similar
to those explored here could be used to estimate the masses of 
bodies residing in and causing structure in other low opacity disks.
%objects such as Beta Pictorus and AU Microscopi that also have disk
%opacity much less than 1.

%We have restricted this initial study of Fomalhaut's ring 
%edge to models with a planet.
%However, alternative models could also be explored.  For example,
%interactions between stellar radiation and dust and gas   
%(e.g., \citealt{klahr01,klahr05} for HR~4796) can
%account for a steep edge surface brightness profile.
%It is possible that a recent collision of two massive
%planetesimals could produce dust via a collisional cascade. 
%Further study is required to determine if these models
%might account for both the
%steepness and eccentricity of Fomalhaut's disk edge. 

A Saturn mass at 119 AU may seem extreme compared to
the properties of our Solar system (Neptune at 30 AU).
It is desirable to place this predicted planet mass in context
with the estimated mass of Fomalhaut's disk.
The total mass required to replenish the dust in the disk was 
estimated by \citet{wyatt02} to be $20-30 M_\oplus$,
however a larger mass is probably required since the velocity dispersion 
assumed by this study corresponded to $h/r \sim 0.1$ and this value exceeds by
a factor of eight that
consistent with the edge slope measured by \citet{kalas05};
$h/r \sim 0.013$.
A power law size distribution with an upper cutoff of 500 km leads to
an estimate of 50-100 Earth masses in the ring \citep{kalas05}.
These estimates suggest that there is sufficient material
currently present in Fomalhaut's disk
to form another Saturn or Neptune sized object.
%Furthermore the disk is quite thin, suggesting that planet formation
%may not have ceased in this system.
%An outer Saturn mass planet is not out of place compared to
%the estimates for the mass in the disk itself.
%
%We note that we have not explored the possibility 
%that Fomalhaut's current morphology and state is transient.
%The current dust opacity could be elevated due
%to a recent collision (e.g., as argued for Vega by \citealt{su05}).
%If so then the mass estimates and the properties of a 
%planet inside the ring would need revision.

% we might predict a kirkwood gap at 144 AU due to the 4:3 resonance
% with width \mu^2/3 = 0.002  x 120 AU = 0.26 AU

%\acknowledgments
\vskip 0.1truein
We thank B. Zuckerman,  R. Edgar,
P. Kalas and E. Ford for interesting discussions.
Support for this work was in part
provided by National Science Foundation grant AST-0406823,
and the National Aeronautics and Space Administration
under Grant No.~NNG04GM12G issued through
the Origins of Solar Systems Program,
and HST-AR-10972 to the Space Telescope Science Institute.

\end{document}